\documentclass[prl,showpacs,twocolumn]{revtex4}  %% REVTeX 4.0
\usepackage{graphics}

\begin{document}

\title{'Shut down' of an atom laser}
\author{N.P. Robins}
\author{A. K. Morrison}
\author{J.J. Hope}
\author{J. D. Close}

\affiliation{Australian Centre for Quantum Atom Optics, The Australian National University, Canberra, 0200, Australia.}
\email{nick.robins@anu.edu.au}
\homepage{http://www.acqao.org/}

\begin{abstract}
We present experimental results demonstrating that the flux of a continuous atom laser beam cannot be arbitrarily increased by increasing the strength of output-coupling.  We find that a state changing output-coupler for a continuous atom laser switches off as the output-coupling strength, parameterized by the Rabi frequency, is increased above a critical value of four times the radial trapping frequency.   In addition, we find that the continuous atom laser has large classical fluctuations when the Rabi frequency is greater than the radial trapping frequency, due to the output-coupling populating all accessible Zeeman states.   The numerical solution of a one-dimensional 5-state Gross-Pitaevskii model for the atom laser is in good qualitative agreement with the experiment.
\end{abstract}


\maketitle

In precision measurement applications, atom lasers have the potential to outperform optical lasers and non-optical techniques by many orders of magnitude provided we can increase their flux, and achieve shot noise limited operation at least in some frequency band.  An outstanding goal in atom optics is to produce a high flux, truly continuous atom laser. In this Letter, we make some early steps along this path and investigate the crossover from weak to strong output-coupling in a continuous atom laser based on a radio-frequency (RF) mechanism.  In this work, we define continuous as extended outcoupling where the width in energy of the outcoupling region is determined by power broadening, rather than pulsed operation, where it is the duration of the pulse that determines the energy width of the outcoupling region. We do not replenish the atoms in the condensate during outcoupling in this work. 

To produce an atom laser, a Bose-Einstein condensate (BEC) is used as a quantum degenerate reservoir of atoms, from which a coherent output coupling mechanism converts atoms from trapped to untrapped states.  Either due to a directed momentum kick \cite{hagley} or gravitational acceleration \cite{mewes,bloch,aspect} the output coupled atoms form a quasi-collimated beam, with the divergence determined by the repulsive interactions between the condensate and atomic beam \cite{aspect}. 
 \begin{figure}[b]
\centerline{\scalebox{.5}{\includegraphics{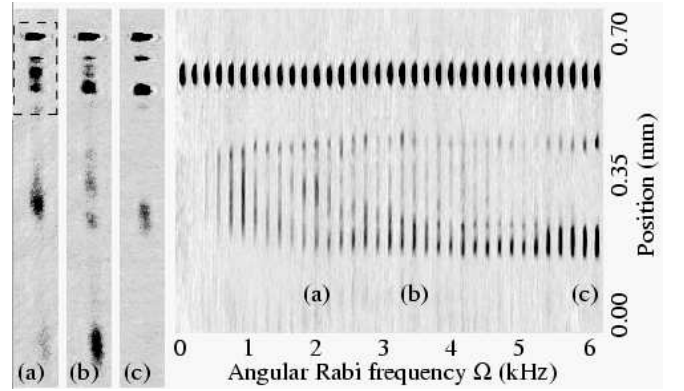}}}
\caption{Optical depth plot (35 independent experiments)  showing the spatial structure of a 3 ms atom laser as a function of output-coupling strength.  The system was left to evolve for a further 4 ms before the trap was switched off and 2 ms later the images were acquired.  The field of view for each individual image is $0.7\times 0.3$ mm.  The final image is taken with $\Omega$=16 kHz.  Anti-trapped Zeeman states  are clearly visible in (a), (b), and (c) which show an extended view of the data in the main part figure, ($2.7\times 0.3$ mm).}
\end{figure}
Our previous experiments on a pulsed output-coupler suggested that a {\em continuous} atom laser would have a stringent limit on peak homogeneous flux \cite{robins}.   Here we show that peak flux into the magnetic field insensitive state is indeed significantly below that which can be provided by the finite reservior of BEC atoms that we produce.  This 'homogeneous flux' limit is imposed by the interaction between multiple internal Zeeman states of the magnetically confined atoms.    Furthermore, we find that a previously predicted effect known as the 'bound state' of an atom laser \cite{hope,jeffers,moy} effectively shuts off state changing output-coupling and hence the atom laser beam.  

The experimental data presented in Figure 1 encapsulate many of the results described herein.  The figure shows the densities of the condensate and atom laser beam in a series of experiments with a 3 ms continuous atom laser, produced in the F=2 manifold of ${\rm ^{87}Rb}$ by state changing output coupling to a magnetic field insensitive state, ${\rm m_F=0}$.    At low output-coupling strength the atom laser beam flux increases gradually and homogeneously until $\Omega\approx$1 kHz, where $\Omega=\frac{g_F \mu_B B}{2\hbar}$ is the bare Rabi frequency.
At around this value we observe that the anti-trapped ${\rm m_F=-2,-1}$ states begin to play a part in the atom laser dynamics.   This leads to increasingly severe fluctuations in the density of the atom laser beam, and a loss of atoms to the anti-trapped states (Fig 1(a),(b)).  A further increase in the coupling strength incrementally shuts down the output starting around $\Omega\approx$ 4 kHz .  In the limit of large output coupling strength (to be defined later) we find approximately 70\% of the atoms remain localized in the condensate, while the other 30\% are emitted shortly after the beginning of the output coupling period.    

The rest of this paper is organized into three parts.  First, we describe the details and relevant time-scales of our experiment.  In the second part we examine, experimentally, the regime of weak to intermediate output-coupling.  In the third part of the paper we demonstrate the 'shut down' or bound state of the atom laser.

\begin{figure}[b]
\centerline{\scalebox{.45}{\includegraphics{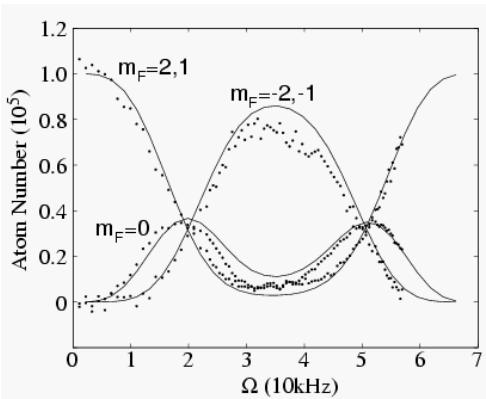}}}
\caption{Outcoupled fraction in the Zeeman states of the F=2 single pulse (${\rm 50\mu s}$) atom laser as a function of angular Rabi frequency.  Solid lines: theoretical calculation with no free parameters from a 3D GP model of the 5-state F=2 atom laser. }
\end{figure}
 In our experiment, we 
produce an ${\rm F=2, m_{F}=2}$  ${\rm ^{87}Rb}$ condensate, consisting of 
approximately $10^5$ atoms, via evaporation in a water-cooled QUIC 
magnetic trap \cite{esslinger} with a radial trapping frequency of
$\nu_\bot=260$ Hz and an axial trapping frequency $\nu_z=20$ Hz. 
We operate our trap at 
a bias field of only $B_0=0.25$ G.  The trap runs at 12 A (12 V)  
generated from a low noise power supply.
The low power dissipation suppresses heating related drift of the magnetic trap bias 
field, allowing precise addressing of the condensate with resonant RF radiation (we measure a drift of significantly less than 0.7 mG over 8 hours).  Unless otherwise stated, in all experiments described in this paper the RF cut is placed 2 kHz from the upper edge of the condensate.
After 50 s of evaporative cooling, the BEC is left 
to equilibriate for 100 ms.  An RF signal generator, 
set in gated burst mode, produces a weak 3-200 ms pulse which is 
radiated perpendicular to the magnetic bias field of the trap through 
a 22 mm radius single wire loop, approximately 18 mm from the BEC.  
The peak RF magnetic field magnitude, B, and hence the bare Rabi frequency, $\Omega$, is calibrated by fitting the frequency axis of a ${\rm 50 \mu s}$ pulsed experiment with a numerical simulation of pulsed output coupling utilizing the 3D Gross-Pitaevskii theory of the atom laser described in our previous paper \cite{robins}.  The results of such a calibration are shown in Figure 2.  We find that the output coupling magnetic field seen by the atoms is reduced by around a factor of 5 compared to a simple calculation of the field radiated from a single loop or a free space measurement of the field from the coil.   This reduction is due to the close proximity of the quadrupole trapping coil on which the RF coil is mounted.  Using this method we estimate the uncertainty in the Rabi frequency to be around 3-5\%.  The quadratic contribution of the Zeeman effect creates resonance shifts of  approximately 0.02\%  at 1 G, and hence our system is modeled accurately by the 5-state GP equation.   
\begin{figure}[t]
\centerline{\scalebox{.35}{\includegraphics{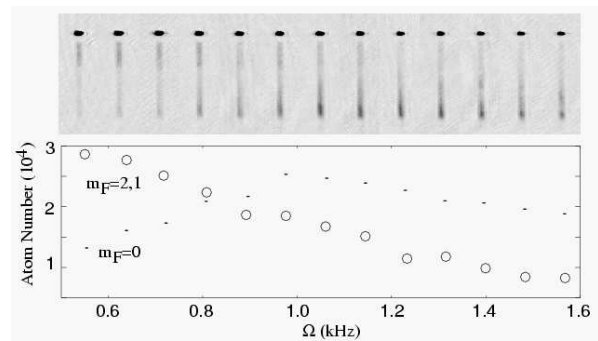}}}
\caption{Atom number as a function of angular Rabi frequency for a 10 ms atom laser.  The upper plot shows the raw Optical depth data averaged over 6 identical experiments.  The field of view for each image is 1.5mm by 0.5mm. }
\end{figure}

To put this work into context, it is critical to understand the time and length scales relevant to the operation of a state changing output-coupler.  In the Thomas-Fermi limit the addressable frequency width of a condensate in the $F=2,m_F=2$ state is given by $W=\frac{g}{\hbar\omega_\bot}\sqrt{2M\mu}$, where $\mu$ is the chemical potential, M is the atomic mass, g is gravity and $\omega_\bot$ is the radial trapping frequency \cite{bloch} (For our condensates $W=2\pi\times 5$ kHz).   For the continuous atom laser the resonance width of the output coupling is determined by the Rabi frequency, $\Omega$, due to power broadening.  A number of definitions exist in the literature for what constitutes weak output-coupling, all based on a comparison of the coupling strength, $\Omega$ with various parameters.   All require that the resonance width of the output coupling is significantly less than the frequency width of a condensate.  The most stringent of these definitions requires $\Omega\ll\omega_\bot\sim1.6$kHz \cite{steck,schneider}.  A more relaxed condition states that $\Omega\ll1.6\sqrt{g/R}\sim3.2$kHz, where $R$ is the spatial Thomas-Fermi width \cite{aspect_theory}.  The least stringent condition is $\Omega\ll1/\tau\sim12$kHz, where $\tau$ is the time it takes atoms to leave the output coupling resonance \cite{graham}.   Strong output coupling can be considered to occur when the weak output-coupling conditions are not met.  
\begin{figure}[b]
\centerline{\scalebox{.6}{\includegraphics{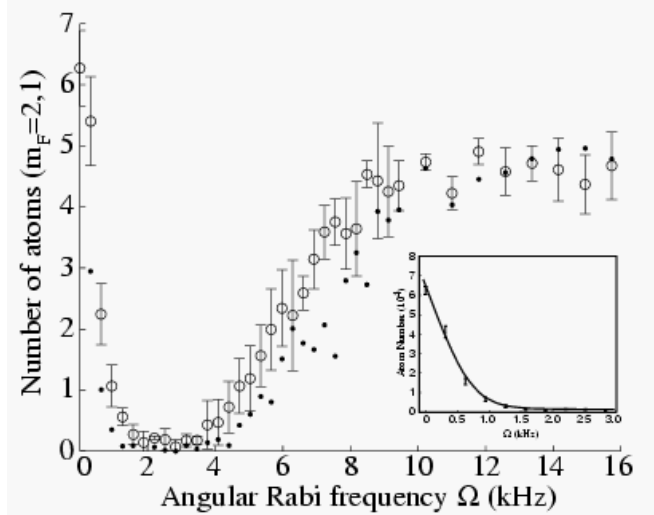}}}
\caption{Trapped fraction as a function of angular Rabi frequency for a 100 ms atom laser (open circles).
Data for a 200 ms output-coupler is shown for comparison (black dots).}
\end{figure}
%A number of these theoretical works have suggested that in the weak limit of the F=2 or F=1 system the anti-trapped magnetic states will not play a part in the dynamics of atom laser output coupling \cite{ss, aspect, busch, japha and on and on}.   These papers make {\em quantitative} comparisons with the experiment of Bloch et. al. \cite{blah} based on this assumption and find good agreement, despite the fact that by their own calculations Bloch's experiment was {\em not} in the weak coupling regime.  In fact, in the experimental work of Bloch et. al. \cite{blah} no discussion of the anti-trapped atomic states was made.  

In Figure 3 we display experimental data taken under a similar set of parameters to Bloch et. al \cite{bloch}.   A 10 ms atom laser beam is produced and allowed to separate from the condensate prior to imaging.  The decline of the trapped condensate atom number clearly follows a simple monotonic decay as found by the previous work \cite{bloch}.  However, from this data it can be seen that the decline of the population in the trapped states does not lead to an increase in the population of the $m_F=0$ state that forms the atom laser beam.   Our conclusion is that the missing atoms have been expelled into the anti-trapped states.  We find the same behavior for all output-coupling times that produce a fully observable atom laser beam (up to around 25 ms for our imaging field of view).  For shorter output-coupling times, such as presented in Figure 1, we are able to observe the anti-trapped states, which appear above an angular Rabi frequency of $\Omega\approx 800$ Hz.  For all our data, we find that the peak output flux occurs at $\Omega\approx 1$ kHz, and that this value corresponds to the most homogeneous atom laser beam.
As $\Omega$ is increased above 1 kHz the atom laser beam develops steadily worsening density fluctuations, the most prominent feature of which is a peak in the leading edge of the beam.    One way around this type of flux limiting dynamics is to reduce the atom laser system to only two levels: trapped and un-trapped.  Such a system has been demonstrated by Aspects group, utilizing the nonlinear Zeeman shift in the F=1 manifold to create an effective two level system.  A Raman transition directly from trapped to un-trapped states with a two photon momentum transfer should also increase the homogeneous flux limit as it removes atoms more quickly from the output-coupling region.
However, in what follows we show that for all continuous state changing output-coupling, including two state systems, output flux is limited by the underlying physics of the process.  
\begin{figure}[t]
\centerline{\scalebox{.65}{\includegraphics{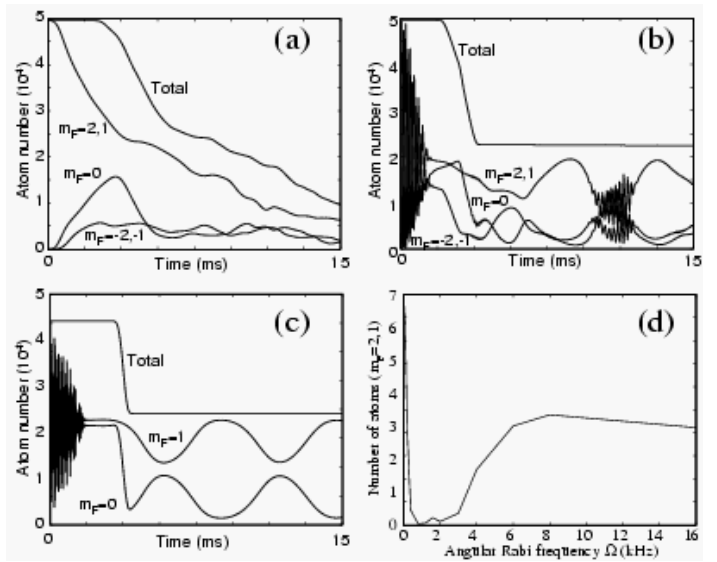}}}
\caption{Theoretical results obtained by solution of the 1D-GP equation. Atomic population in the Zeeman states as a function of time for a 15 ms atom laser for (a) $\Omega=800$ Hz and (b) $\Omega=16$ kHz.  In (c) we show the behaviour of a 'pure' two state atom laser system for $\Omega=16$ kHz.  (d) Total atomic population remaining in the magnetic trap after 100 ms of output coupling plotted as a function of angular Rabi frequency.}
\end{figure}

In Figure 4. we display a measurement of the number of atoms remaining in the trapped states as the Rabi frequency is scanned from the weak to strong output-coupling regime for a 100 ms continuous atom laser (open circles in the figure). 
Note that for this output coupling time the peak flux in the atom laser beam is below our detection sensitivity.
The condensate is progressively depleted at increasing, but weak, Rabi frequencies (Figure 4, inset) until we observe no atoms remaining in the magnetic trap.  At $\Omega\sim 4$ kHz atoms start to reappear, and by $\Omega\sim 9$ kHz the number of atoms remaining in the trap stabilizes to 70\%
of the initial condensate number.  As described earlier, the other 30\% are expelled at the initial switch on of the output-coupling.     Our method of collecting data ensures that there are no systematic shifts in the trap bias field.  Every 10 runs of the experiment we switch to a 10 ms atom laser to ensure constant detuning, as well as comparing detailed data to a 'coarse' set taken at the beginning of each run. 
We emphasize here that the Rabi frequencies over which this effect occurs correspond to the cross-over from the weak to strong coupling regime.  In order to measure atoms remaining in the BEC after 100 ms  of output-coupling, the combination of the magnetic trap and RF field in the cross-over regime must be creating a trap for the atoms.  Equivalently, the atom laser ceases to operate, or 'shuts down' in the intermediate to strong coupling regime.     We have verified that the 'shut down' edge is independent of output coupling duration, as shown in Figure 4 with a 200 ms atom laser (black dots).   For shorter coupling times the condensate is not fully depleted before this edge is reached (see for example Figure 1).

A number of theoretical works have suggested that the atom laser may have a 'bound'  eigenstate \cite{hope,jeffers,moy}, based purely on the existence of coupling between a single trap mode and a continuum of un-trapped states.  Furthermore, in the context of producing a two dimensional BEC, it has been shown recently that trapping of {\em all} $m_F$ states is a natural consequence of combining RF coupling with a DC magnetic trap \cite{zobay}.  This trapping can be understood by considering the 'dressed state' basis in which the RF coupling and DC potentials seen by the atoms are diagonalised.   In this basis the dressed eigenstates are linear combinations of the bare Zeeman states, trapped in effective potentials created by the avoided crossings.    In the strong coupling limit, for the F=2 atom laser, diagonalization yields a prediction of between 31.25\% up to a maximum of 62.5\% of the atoms remaining trapped, assuming a non-adiabatic projection onto the dressed states.    Our experimental observation of 70\% is inconsistent with these values, suggesting some degree of adiabatic transfer.   In order to gain further qualitative insight we numerically solve the F=2 Gross Pitaevskii model \cite{robins} of the atom laser in one dimension given by 
\begin{equation}
 \label{secondset}
\begin{array}{l} 
{\displaystyle
i\dot\phi_2=({\mathcal{L}}+
z^2+Gz-2\Delta )\phi_2+2\Omega\phi_1}\\[7pt]
{\displaystyle
i\dot\phi_1=({\mathcal{L}}+
\frac{1}{2}z^2+Gz-\Delta )\phi_1+2\Omega\phi_2+\sqrt{6}\Omega\phi_0}\\[7pt]
{\displaystyle
i\dot\phi_0=({\mathcal{L}}+Gz) \phi_0+\sqrt{6}\Omega\phi_1+\sqrt{6}\Omega\phi_{-1}}\\[7pt]
{\displaystyle
i\dot\phi_{-1}=({\mathcal{L}}
-\frac{1}{2}z^2+Gz+\Delta )\phi_{-1}+2\Omega\phi_{-2}+\sqrt{6}\Omega\phi_0}\\[7pt]
{\displaystyle
i\dot\phi_{-2}=({\mathcal{L}}-
z^2+Gz+2\Delta )\phi_{-2}+2\Omega\phi_{-1} \,  ,}
\end{array}\end{equation}
where $\phi_i$ is the GP function for the $i$th Zeeman state and 
${\mathcal{L}}\equiv-\frac{1}{2}\frac{\partial^2}{\partial x^2}+U (\Sigma_{i=-2}^{2} |\phi_{i} |^2)$. 
Here $\Delta$ and $\Omega$ are respectively the 
detuning of the RF field from the $B_0$ resonance, and the Rabi frequency, measured in units of the radial trapping frequency $\omega_\bot $ (for the $m_F=1$ state), $U$ is the interaction coefficient and $G=\frac{mg}{\hbar 
\omega_\bot }(\frac{\hbar}{m\omega_\bot })^{1/2}$ gravity. The wave functions, time, spatial coordinates, and interaction strengths are 
measured in the units of $(\hbar/m\omega_z)^{-1/4}$, $\omega_\bot ^{-1}$, $(\hbar/m\omega_\bot )^{1/2}$, and $(\hbar \omega_\bot )^{-1}(\hbar/m\omega_\bot )^{-1/2}$, respectively.  The nonlinear interaction strength is derived by requiring that
the 1D TF chemical potential be equivalent to the 3D case.  There are no free parameters in this model; we use  U=$6.6\times10^{-4}$, G=9.24, $\Omega=0-13.85$, $\Delta=8.2$.

Figure 5(a) shows the time evolution of the population numbers in the trapped, un-trapped and anti-trapped Zeeman states for $\Omega=800$ Hz.  The trapped $m_F=2,1$ populations tend towards zero, essentially monotonically.
In 5(b), corresponding to $\Omega=16$ kHz, we find an entirely different behavior.  After an initial high frequency exchange of the number of atoms in the different Zeeman states, slightly more than 50\% of the atoms are ejected from the system.  The remaining atoms, which populate {\em all} Zeeman states, undergo a lossless and slow exchange of population.  We identify the trapped atoms as being in a superposition of the dressed state eigenvectors.  The ejected atoms simply correspond to the un(anti)-trapped dressed states.  5(c) shows the same trapping effect, but this time in a simple two-level atom laser system, indicating that the bound mode is inherent to state changing output-coupling regardless of the number of states involved.   Finally, 5(d) shows a prediction of the number of atoms remaining in the trap after 100 ms of output coupling as a function of Rabi frequency.  The behaviour qualitatively matches that of the experiment, indicating that we are indeed observing the 'bound state' of the atom laser in our experiments.   The predicted number of atoms remaining in the trap lies within the range of the strong coupling estimate, but still differs substantially from our experimental results.  We expect that a full 3D model would capture the dynamics of our experiment, but it is beyond the scope of the present work.

In conclusion, we have shown that the peak homogeneous output flux of an atom laser beam derived from a finite BEC is limited by the requirement to operate well within the weak coupling regime, in order to avoid pumping to anti-trapped states and the related density fluctuations.  An obvious short term solution to boost the flux would be to chirp the RF detuning across the condensate, however such a scheme would not be applicable to a pumped continuous atom laser.   Additionally, we find that one cannot arbitrarily increase the output flux of a state changing atom laser due to the presence of bound modes.  We are currently investigating momentum kicked continuous Raman coupling and condensate pumping as a means of boosting atom laser flux.
\acknowledgments
{NPR thanks S.A. Haine, C. M. Savage, C. Figl and E.A Ostrovskaya for the many discussions related to this work.}

\end{document}